# 3D Interconnected Magnetic Nanowire Networks as Potential Integrated Multistate Memristors

Dhritiman Bhattacharya[1], Zhijie Chen[1], Christopher J. Jensen[1], Chen Liu[2], Edward C. Burks[3], Dustin A. Gilbert[4], Xixiang Zhang[2], Gen Yin[1], and Kai Liu[1*]

[1]Physics Department, Georgetown University, Washington, DC 20057, USA

[2]Physical Science and Engineering Division, King Abdullah University of Science & Technology, Thuwal 23955-6900, Saudi Arabia

[3]Physics Department, University of California, Davis, CA 95618, USA

[4]Department of Materials Science and Engineering, and Department of Physics and Astronomy, University of Tennessee, Knoxville, TN 37996, USA

## ABSTRACT

Interconnected magnetic nanowire (NW) networks offer a promising platform for 3-dimensional (3D) information storage and integrated neuromorphic computing. Here we report discrete propagation of magnetic states in interconnected Co nanowire networks driven by magnetic field and current, manifested in distinct magnetoresistance (MR) features. In these networks, when only a few interconnected NWs were measured, multiple MR kinks and local minima were observed, including a significant minimum at a positive field during the descending field sweep. Micromagnetic simulations showed that this unusual feature was due to domain wall (DW) pinning at the NW intersections, which was confirmed by off-axis electron holography imaging. In a complex network with many intersections, sequential switching of nanowire sections separated by interconnects was observed, along with stochastic characteristics. The pinning/depinning of the DWs can be further controlled by the driving current density. These results illustrate the promise of such interconnected networks as integrated multistate memristors.

Finding a hardware platform compatible with artificial intelligence and neuromorphic architectures is a key challenge of future computing. Three-dimensional (3D) systems are well suited to address this need, as harnessing the complex nature and degrees of freedom associated with their structure can provide enriched functionalities. 3D nanomagnetic devices are particularly promising due to their energy efficiency, non-volatility, and scalability,[1-9] and can be utilized to implement sophisticated racetrack magnetic memory,[10-12] logic[13-14] and neuromorphic computing.[15-16] Recent advent of novel nanofabrication techniques[17-20] has led to rapid progress in complex 3D magnetic nanostructures such as curved or flexible thin films,[2, 4] thickness-modulated nanowires,[21] magnetic nanospirals[22] or nanohelices,[23] etc. Such assemblies exhibit unique topological features and can accommodate exotic spin textures, with wide-ranging potential applications in spintronics,[1-5, 7, 23-24] magnonics,[25] and artificial spin ice.[26]

To be integrated into neuromorphic circuitry, it is crucial to realize a 3D nanomagnetic system hosting many different magnetic states that can be controlled by certain external stimuli. In this regard, only a few experimental options have been explored to date and mostly limited to 2D arrays,[16, 26-31] as it is highly challenging to achieve 3D systems with desired characteristics. We previously identified interconnected magnetic nanowire networks as a promising platform for 3D information storage and neuromorphics,[19] where unique magnetization reversal mechanisms through intersection-mediated domain wall pinning and propagation were observed. These results suggest that it might be possible to encode digital information into the magnetic state of the networks, and propagate it through the interconnected magnetic networks to implement a network of repeatable multi-state memristors in neuromorphic circuits.[16, 32]

To demonstrate this feasibility, here we have employed electrical transport studies, together with magnetometry, magnetic imaging and micromagnetic simulations to probe magnetic characteristics and switching pathways in the complex nanowire networks. We report discrete propagation of magnetic states in interconnected Co nanowire networks through magnetoresistance (MR) measurements under different applied field geometry, strength and driving current density. As a result, information encoded in the spin texture can propagate in the networks from one section to another. These findings demonstrate the potentials for 3D information storage and integrated multistate memristors.

The Co networks were synthesized by ion tracking of 6 μm thick polycarbonate membranes at multiple azimuthal angles with equal fluence, followed by electrodeposition of cobalt nanowires, as described earlier.[19] The ion tracking was performed at 3 angles, normal to the membrane and at 45° colatitude angle, the latter with 0° and 180° azimuthal orientations, respectively. These three tracking directions defined an ion-tracking plane, as illustrated in Fig. 1(a). For this study, networks of 75nm diameter Co nanowires (NWs) with $10^9$ intersections / cm$^2$ were used.[19] Scanning electron microscopy (SEM) imaging revealed that the nanowire networks were quasi-ordered with interconnects mostly lying in the ion-tracking plane (Figs. 1a and 1c), while some nanowires remained isolated. Fast Fourier transform (FFT) of the image indicated the presence of NWs at three different angles (Fig. 1a inset). The location of intersection could be anywhere along the length of the wire, and the depth of intersection could vary from just touching to completely going through each other. The nature of intersection (i.e. depth and location) becomes important to account for the MR characteristics discussed later.

Magnetic properties were measured by vibrating sample magnetometry (VSM) at room temperature with external field up to 1.2 T with NWs inside the membrane. The measurements were performed with the field applied at an angle φ relative to the ion tracking plane normal [Fig. 2(a)]. An almost square loop was found when the field was perpendicular to the polycarbonate membrane (φ = 90°, parallel to the ion-tracking plane yz), indicating a magnetic easy axis. On the other hand, more slanted loops with smaller remanence and coercivity were observed when the field was perpendicular to the ion-tracking plane (φ=0°). The effect of individual intersections is not prominent in these magnetometry measurements.

Previously, MR in cylindrical nanowires have been extensively investigated and ordinary, anisotropic and giant magnetoresistance have been identified in different material systems.[33-35] MR studies in connected 2D nanowire networks exhibited significant contributions from the vertices where the NW segments meet.[36-37] In complex 3D magnetic nanostructures, additional unusual angular dependence of MR may emerge, as illustrated recently in nanobridges.[8]

In the present study, MR measurements were performed on over 50 samples of interconnected nanowire networks inside polycarbonate membranes. Details of the measurement procedure are provided in the Supporting Information. The MR behavior in the network originates from the anisotropic magnetoresistance (AMR) of Co. Conventionally, AMR depends on the relative alignment between the magnetization and the current flow: $R = R_T + \Delta R cos^2\alpha$, where $\alpha$ is the angle between the magnetization and the current, and $R_T$ is the resistance when the magnetization and the current are perpendicular to each other. As the current flows along the long axis of the NW; magnetization parallel (perpendicular) to the current would yield a high (low) resistance.

First, a collection of nanowires was measured, with a resistance of ~3Ω [Fig. 2(b)]. This is roughly equivalent to $10^2$ pairs of intersecting NWs in parallel (Supporting Information). The MR was defined as [R($H$) - $R_{max}$]/$R_{max}$ (%), where R($H$) and $R_{max}$ represented the resistance in an applied field $H$ and the maximum resistance along a MR curve, respectively. When the field was applied perpendicular to the ion-tracking plane ($\varphi$ =0°), the magnetization gradually rotated from a positively saturated state to an intermediate state with magnetization aligned within the ion-tracking plane, and finally to the negative saturation direction during a descending-field sweep. The process was mostly reversible in an ascending-field sweep, leading to a bell-shaped MR curve [Fig. 2(b) bottom panel]. When the field was applied within the ion-tracking plane ($\varphi$ =90°), a sharp resistance minimum was found near the negative coercive field during the descending-field sweep, along with a symmetric minimum near the positive coercive field during the ascending-field sweep [Fig. 2(b) top panel]. For the $\varphi$ =45° case, a qualitatively similar AMR behavior was observed as that in $\varphi$ =90°. These are typical AMR characteristics, similar to those in non-intersecting, parallel nanowire arrays.[38] As a collection of NWs was measured, characteristics of individual interconnected nanowires, even if different, were likely averaged out.

Second, fewer nanowires were connected for MR measurements by selecting regions with fewer overgrown NWs. One such example is presented in Fig. 2(c). The resistance value of 200Ω corresponded to roughly 70-fold fewer NWs compared to the previous case. Indeed, the effect of intersection became more evident here. In the $\varphi$=0° case, a sharp discontinuous MR jump was observed, corresponding to irreversible DW switching/realignment. In the $\varphi$= 45° and 90° cases, *multiple* MR kinks were observed, indicative of domain wall pinning and depinning in the nanowires. This is more clearly illustrated in Fig. 2(d) for $\varphi$= 90° (blue curve), while the aforementioned low resistance case exhibited one minimum near coercive field (green curve). Furthermore, for a descending field sweep from positive saturation, *a significant MR minimum is found at a positive field*, in contrast to the usual case with a minimum in a negative field. This unusual minimum can be attributed to the formation of DWs at the intersection [39]. The overall MR behavior can be ascribed to DW formation and pinning at the intersection and subsequent realignment of magnetization in a section of the NW networks via DW depinning. That is, the magnetization reversal mechanism can be characterized as step-by-step switching of separate sections of the NWs demarcated by the intersections. We have also corroborated this by performing 3D micromagnetic simulations using MuMax3[40] (Supporting Information). As mentioned earlier, the location of intersections along the nanowire axis and depths of overlap vary in different parts of the networks [Fig. 1(b, c)]. Consequently, DW nucleation and depinning fields are also expected to vary among different intersections. Few such cases are illustrated in Fig. S3 of the Supporting Information, highlighting the influence of interconnect geometry on magnetization reversal.

So far, MR measurements are presented for two different cases: (i) many nanowires, which exhibit low resistance with one minimum during descending-field sweep; (ii) few interconnected NWs, which exhibit high resistance with multiple minima. In this section, we present MR measurements for a third sample region with low resistance (~4Ω) and multiple discrete switching steps, corresponding to many NWs with a high density of interconnects. MR measurements performed at different applied magnetic field angle φ are summarized in Fig. 3. For φ=0°, a bell-shaped MR curve similar to that shown in Fig. 2(b) was found, corresponding to reversible magnetization switching. When the magnetic field was slightly tilted (φ =10°), while the primary MR feature still remained mostly the same at high fields, small MR peaks appeared at low fields, indicative of switching of different NW sections. At φ =25°, the ascending and descending MR branches began to diverge noticeably, forming discrete and hysteretic MR jumps (highlighted by ovals). As φ was increased further, these jumps became sharper (marked by triangles) and consequently the hysteresis became more pronounced (φ =38°-45°). Similar sharp MR features were previously observed in 2D artificial spin ice structures, particularly in connected planar nanowire networks, due to collective switching of a group of nanowires governed by the ice-rule, i.e. the symmetry of the system.[37, 41-42] In our system, the sharp MR jumps are manifestations of the intersections, and could be attributed to switching of a subset of the densely connected network of many nanowires. Different sections switched at different times due to the variation of local energy landscape as the number of intersections, the locations of intersections along the nanowire axis and the depths of overlap varied across the network. For the same reason, the number of switching steps depended on the relative angle between the NWs and the magnetic field. For example, as many as five discrete jumps were observed for φ=42°. The switching steps also evolved in a consistent manner with the magnetic field angle as can be realized by tracking different switching events S1-S5 for φ =38°-45° (Fig. 3 middle column, marked by triangles). Here, with increasing φ, S3 moved progressively towards zero field, while S1, S2, S5 moved towards larger reversal fields.

Finally, the possibility of controlling DW nucleation and pinning/depinning at the intersection using current pulses was explored in a network with low resistance (~3Ω) and few different switching steps. With pulses of appropriate magnitude, the peak position (i.e. DW depinning field) monotonically changed [Fig. S4(a), φ=35°]. Similar trend was observed when the tilting angle was changed to φ=90° albeit with a different slope of DW depinning field vs. applied current [Fig. S4(b)]. Several phenomena such as current induced spin torque, Oersted field or Joule heating could contribute to this behavior [43-47]. This controllable stabilization and propagation of magnetic states enable one to selectively address different subsections of the network by tuning the strength and orientation of magnetic field as well as current pulses.

The abundance of magnetic states that can be stabilized in the network also leads to the possibility of addressing the network by applying different field sequences. To closely investigate the effect of history of prior magnetic state on the reversal behavior, we employed the first-order reversal curve (FORC) technique.[48-51] Here, the sample location was the same as mentioned in the previous section with a tilting angle φ =35°. The major loop is shown in Fig. 4(a). MR-FORCs were measured by the following procedures.[36, 52] After positive saturation the applied field was reduced to a given reversal field $H_R$. From this reversal field (i.e. a point on the descending branch of the major loop), the resistance was then measured as the applied field $H$ was swept back towards positive saturation, thereby tracing out a single MR-FORC. This process was repeated for a series of decreasing reversal fields, creating a family of FORCs. An animation of the full FORC family is provided in the Supporting Information. The FORC distribution was then defined as a mixed second-order derivative of the measured MR:

$$FORC\ Distribution = -\frac{1}{2}\frac{\partial^2 MR(H,H_R)}{\partial H\ \partial H_R} \quad (1)$$

Hence, only the irreversible switching events are mapped out in the FORC distribution,[48, 51] as can be seen by the highlighted FORC features in Fig. 4(b).

For the primary FORC feature centered at ($H$~ 0.55 kOe, $H_R$ ~ -0.40 kOe), highlighted by the horizontal red oval in Fig. 4(b), the corresponding individual FORCs are illustrated in Fig. 4(c), with -1.2 kOe < $H_R$ < 0, spanning over the largest MR minimum of the major loop centered around $H$= 0.37 kOe. The MR-FORC slope decreases (increases) to the left (right) of $H$= 0.37 kOe at successively more negative $H_R$ values, leading to a negative-positive pair of FORC features.[48, 51]

Besides this primary minima, four discrete MR jumps were also present in the φ=35° case, marked as S2-S5 in Fig. 4(a) along the ascending-field sweep (red curve). These discrete MR jumps can be further analyzed by correlating the FORC features with individual FORCs. For example, as shown in Fig. 4(d), switching events S2, S3 only took place when the reversal field $H_R$ < -3.5 kOe, marked by the black line in Fig. 4(d), which itself corresponds to a switching event S5 along the descending branch. That is, only when a portion of the networks switched (corresponding to S5 along the *descending-field* branch, gray curve) would subsequent switching events occur (corresponding to S2 and S3 along the *ascending-field* branch, red curve). *This trend clearly illustrates the hysteretic nature of the switching steps and their dependence on the prior magnetic state.* The corresponding negative and positive FORC features are marked as S2 and S3 in Fig. 4(b). Similarly, a number of other switching events, highlighted by unlabeled, dashed rings in Fig. 4(b), can be correlated to the magnetic state of the network initialized during the FORC measurement procedure to a particular $H_R$ starting from saturation. Therefore, *reversal of a part of the network allows access to new magnetization reversal pathways and consequently determines the final magnetic state at any given field.* Thus, the network can be conditioned to host different magnetic states by varying the sequence of applied magnetic field. These MR-FORC characteristics are reminiscent of the return point and complementary point memory effects induced by disorders in certain magnetic systems, where the microscopic magnetic configurations are correlated with prior magnetic field history, or retaining certain microscopic "memory".[53-54] In the present system, the network structure plays the same role as disorders in anchoring the energy landscape during magnetization reversal to guide the system to a particular magnetic state.

A particularly noteworthy negative FORC feature was found in the vicinity of S5 in the FORC distribution (Fig. 4b inset). The total FORC distribution was calculated by averaging five individual FORCs measured at each reversal field (Supporting Information). These individual FORCs were examined to find the origin of this feature. We note that most of the MR jumps are reproducible, as they occur in a definitive fashion after a certain applied magnetic field sequence. However, in the range -6.8 kOe $< H_R <$ -5.5 kOe, for the same reversal field $H_R$, irreversible switching steps appeared at different applied fields $H$. Examples for two different reversal fields are presented in Fig. 4(e). Here, the green curves correspond to $H_R=$ -5.8 kOe. Although the reversal field is the same, one FORC contains a switching step at $H=$2.5 kOe while the other at $H=$ 3.5 kOe. These regions are respectively shaded by black and red boxes to highlight the MR jumps. A similar scenario is shown for $H_R=$ -6.3 kOe by the blue curves. Therefore, even though the network was initialized by a definitive magnetic field sequence to a specific $H_R$ value, the actual magnetic state for 2.5 kOe $< H <$ 3.5 kOe was randomly determined. This could happen if there are multiple switching pathways separated by an energy barrier comparable to the thermal energy $k_BT$.

To visualize the magnetic configurations of the nanowires, especially at the intersections, magnetic imaging was performed using off-axis electron holography. This technique directly measures the phase shift of the electron wave that passes through the specimen. Then the quantitative information about the electrostatic potential and the in-plane component of magnetic induction within and surrounding the specimen can be separated in the total phase shift by a flipping method (Supporting Information). Firstly, imaging of an isolated wire at remanence was performed. The amplitude and the magnetic contour are shown in Figs. 5(a) and 5(b), respectively. In this case, the induction field lines corresponding to the magnetic isophase contours were found to be parallel to the wire axis. This means a single domain state with magnetization uniformly aligned along the NW long axis was stabilized due to the shape anisotropy of the wire. On the other hand, a multidomain state with a domain wall situated at the intersection was observed at remanence in the interconnected nanowires, pointing to a different direction than the moments along the wires [Fig. 5(c, d)]. This shows the propensity of the DWs to be pinned at the intersection, as shown in the simulated magnetic configuration where one of the wires switched (Fig. S2v). In other words, the intersections are the energy minima for the DWs. It is conceivable that dislodging a DW would require additional energy once it reaches this minimum, which ultimately manifested as discrete jumps in the MR measurements. Hence, the imaging provides further support to our proposed reversal mechanism to explain the MR behavior.

In summary, we report discrete propagation of magnetic states due to DW pinning and depinning in interconnected Co nanowire networks, manifested in distinct MR jumps. This characteristic could be utilized to encode digital information for 3D information storage and integrated neuromorphic computing. Additionally, the pinning / depinning behavior was found to be strongly dependent on the interconnect geometry. These unique fingerprints may be used as challenge response pair (i.e. digital keys) in physically unclonable functions (PUFs), a hardware security primitive based on intrinsic random properties embedded in the physical structure of a device.[55-56]. The MR-FORC measurements emphasized the strong dependence of magnetization reversal on the initial magnetic state or the history. This could be utilized to fashion non-Boolean computing devices such as spintronic memristors and synaptic devices where different resistance states or synaptic weights can be programmed by controllably switching a certain subsection of the networks.[16, 57] Another potential application avenue is reservoir computing, which is a neural network with fixed recurrent weights and thus incur low training cost.[15] The hysteretic behavior and numerous achievable states could allow the interconnected networks to afford large expressivity, a critical reservoir computing feature. Furthermore, the networks exhibit stochastic behavior which is amenable for performing probabilistic computation.[16, 58] We finally showed current pulses can be used as an additional handle to control the magnetic states in the networks.

Thus, this study highlights the potential of interconnected nanowire networks to be used as efficient memristors utilizing their capability to stabilize different magnetic configurations, along with controlled and discrete propagation through the networks. Further investigation of this promising system could ultimately lead to 3D magnetic nanostructure-based energy efficient devices capable of realizing a multitude of functionalities.

## ASSOCIATED CONTENT

**Supporting Information:** The Supporting Information is available free of charge.

Materials and methods, micromagnetic simulations of intersection-mediated domain wall propagation, examples of magnetoresistance measurement on few interconnected nanowires, current driven domain wall propagation through the networks, extended MR data in a network with many nanowires and high density of interconnects, and animation of the MR-FORCs.

## AUTHOR INFORMATION

**Author contributions:** D.B. and K.L. conceived the study and coordinated the project. G.Y. proposed the memristor study. E.B. prepared the nanowire networks. D.B. carried out the MR and magnetometry measurements, as well as the SEM imaging. Z.J.C. performed the micromagnetic simulations. C.L. D.B. and X.X.Z prepared the isolated NW samples and carried out off-axis holography imaging. D.B., Z.J.C., C.J.J., G.Y., D.A.G., C.L., X.X.Z. and K.L. analyzed the results. D.B. and K.L prepared the first draft of the manuscript. All authors contributed to manuscript revision.
**Notes:** The authors declare no competing financial interest.

## ACKNOWLEDGEMENTS

This work has been supported in part by the NSF (ECCS-1933527, ECCS- 2151809), by SMART (2018-NE-2861), one of seven centers of nCORE, a Semiconductor Research Corporation program, sponsored by the National Institute of Standards and Technology (NIST), and by KAUST (OSR-2019-CRG8-4081). We thank Drs. Lisa Debeer-Schmitt, Alexander Grutter, and Julie Borchers for technical assistance with neutron scattering experiments, and Drs. Hajo Frerichs, Lukas Stuehn, and Chris Schwalb for help with magnetic imaging investigations.

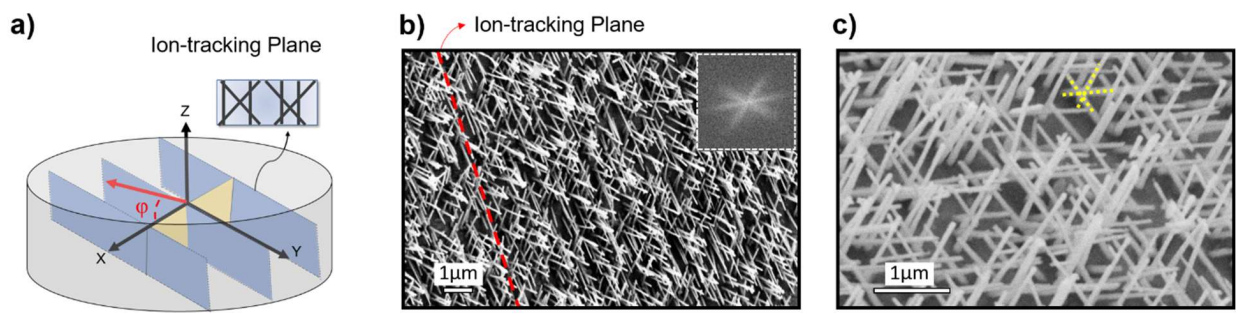

**Figure 1. Network geometry.** (**a**) Schematic of the nanowire networks illustrating the geometry. The ion-tracking plane is shown by the blue box, while the nanowires inside are shown by black lines. φ is the angle of the applied field relative to the ion-tracking plane normal. (**b**) Top view SEM image of the network. The dashed red line shows the ion tracking plane. The inset shows FFT of the SEM image indicating the presence of NWs at three different angles. (**c**) A zoomed-in view of the network. The yellow lines highlight nanowires at three different angles.

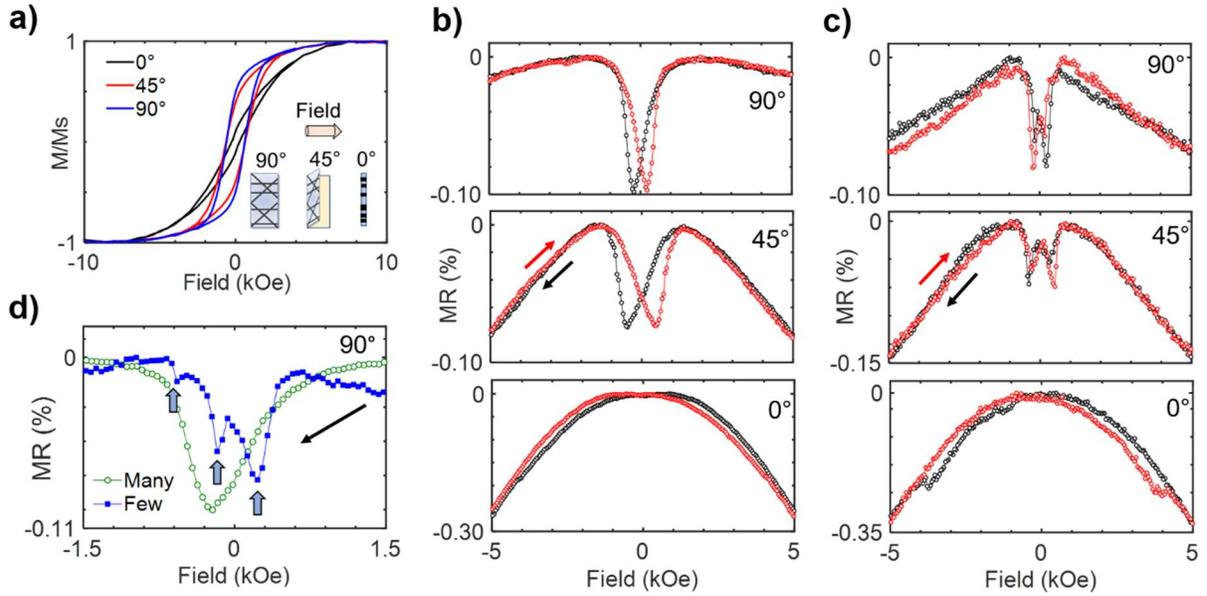

**Figure 2. Magnetometry and magnetoresistance measurements of the NW networks.** (**a**) Magnetometry of the network with the field applied at an angle φ relative to the ion-tracking plane normal. (**b**) MR of a collection of nanowires, R~3Ω. (**c**) MR of a few intersecting nanowires, R~200Ω. (**d**) Comparison of descending-field sweep of MR in (b, c) at φ=90° with many wires (green) and few wires (blue), respectively, showing several pinning and depinning events present in the latter (marked by arrows).

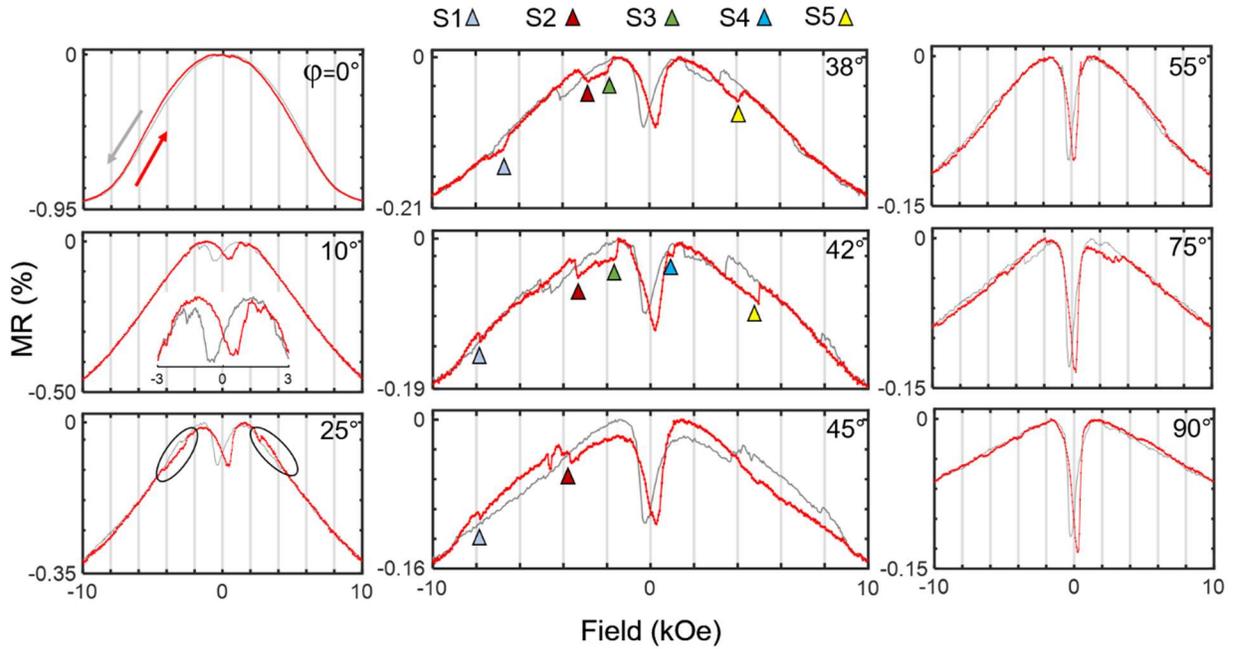

**Figure 3. MR behavior in a network with many nanowires and high density of interconnects.** MR curves are shown for different applied magnetic field angle φ relative to the ion-tracking plane normal, in red/gray for ascending/descending-field sweep, respectively. The hysteresis at φ =25° is shown by the ovals. The triangles mark the different MR jumps for φ =38°-45° along the ascending field sweep (red).

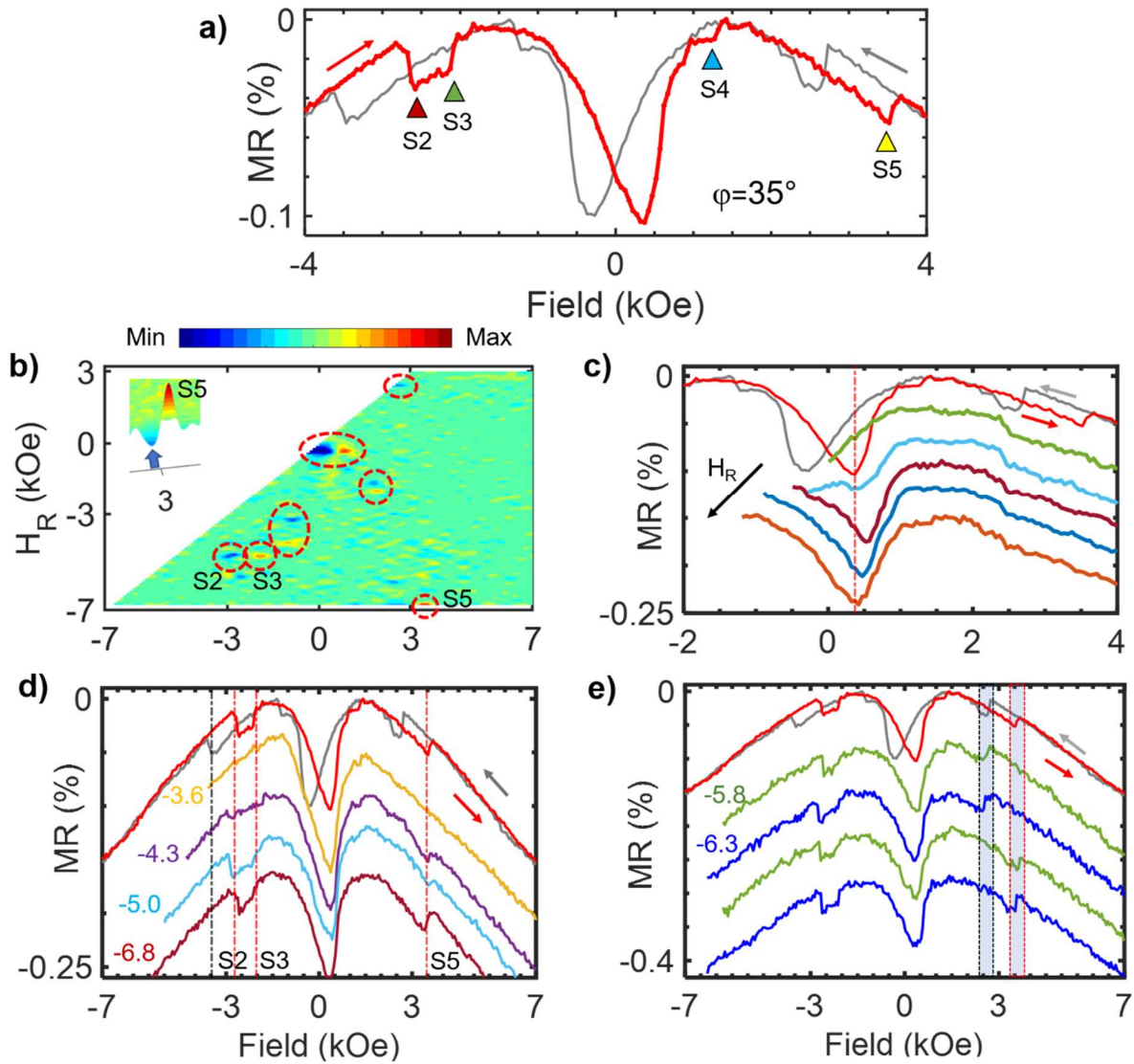

**Figure 4. Magnetoresistance FORC of the Co networks measured at φ=35°.** (**a**) Full-range MR curve (major loop). Arrows mark the sweep direction and triangles mark the MR jumps along the ascending-field sweep (red curve). Symmetric MR jumps along the descending-field sweep are not marked (gray curve). (**b**) FORC distribution plotted in (H, $H_R$) coordinates. The red ovals highlight irreversible switching events. The inset shows a zoom-in view around a FORC feature S5. Representative MR-FORCs around the switching event (**c**) at H= 0.37 kOe, the largest MR minimum, and (**d**) around S2, S3 and S5. (**e**) Stochastic switching behavior at different reversal fields. The green and blue curves correspond to $H_R$=-5.8 kOe and -6.3 kOe, respectively. The switching randomly took place at two different fields highlighted by the shaded area. For (c, d, e) the MR curve is shown on top as a reference.

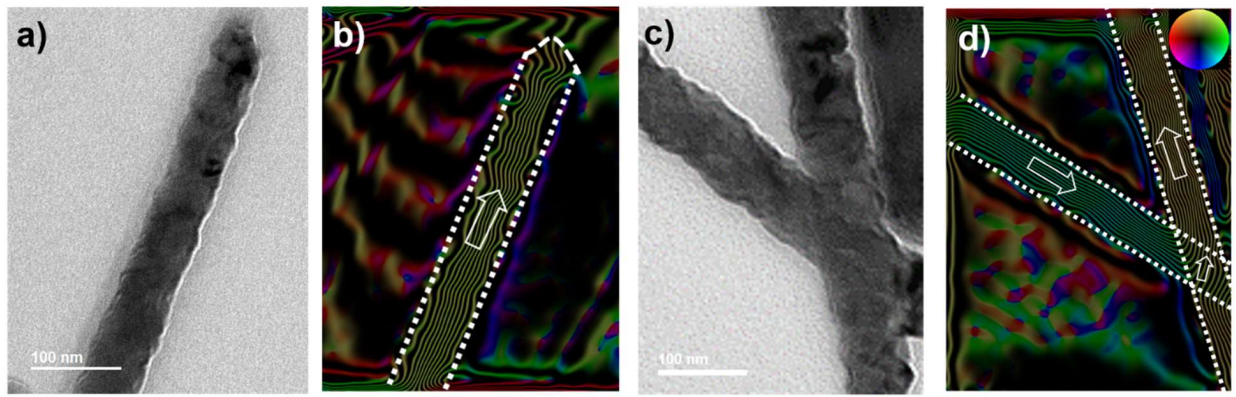

**Figure 5. Off-axis electron holography imaging.** (**a**) Amplitude and (**b**) magnetic contour of an isolated nanowire. (**c**) Amplitude and (**d**) magnetic contour of two interconnected nanowires. The color wheel represents the in-plane direction of the magnetic induction. At remanence, a uniformly magnetized state was observed in the isolated nanowire while a multidomain state with a domain wall at the intersection was stabilized in the interconnected nanowires. The arrows point to the general direction of the magnetization in the nanowires.

Supporting Information

# 3D Interconnected Magnetic Nanowire Networks as Potential Integrated Multistate Memristors

Dhritiman Bhattacharya[1], Zhijie Chen[1], Christopher J. Jensen[1], Chen Liu[2], Edward C. Burks[3], Dustin A. Gilbert[4], Xixiang Zhang[2], Gen Yin[1], and Kai Liu[1]

[1]Physics Department, Georgetown University, Washington, DC 20057, USA
[2]Physical Science and Engineering Division, King Abdullah University of Science & Technology, Thuwal 23955-6900, Saudi Arabia
[3]Physics Department, University of California, Davis, CA 95618, USA
[4]Department of Materials Science and Engineering, and Department of Physics and Astronomy, University of Tennessee, Knoxville, TN 37996, USA

## MATERIALS AND METHODS

### SEM imaging

The interconnected nanowire (NW) networks were fabricated by electrodeposition of nanowires in a polycarbonate membrane with a Cu layer on one side serving as the working electrode.[1] Before performing scanning electron microscopy (SEM), the polycarbonate membrane was partially dissolved by submerging the sample in dichloromethane for a few minutes. The resulting free-standing networks on Cu were examined by a ZEISS-SUPRA A55 SEM at 10 KV accelerating voltage and 20 μm aperture diameter.

### Magnetoresistance measurement

Magnetoresistance (MR) measurements were performed with NWs inside the membrane by applying current pulses using a Keithley 6221 source and reading the voltage using a Keithley 2182 nanovoltmeter at 0.05 kOe field interval. Electrical leads were attached to opposite sides of the membrane using silver paste. The backing electrode on one side of the membrane served as the bottom electrode while the overgrown NWs on the other side of the membrane were used as the top electrode. As the density of overgrown NWs varied across the membrane, different number of NWs could be connected for measurement by changing the top electrode location. In total more

**Supporting Information**

than 50 samples were measured, with each type repeated multiple times. It is hard to determine the exact number of NWs as the nature of the intersections varies across the sample. Therefore, the NW number provided in the main text is only a rough estimate. The resistivity of Co NWs reported previously is in the range of 10-20 µm·cm.[2-3] Considering a resistivity of 15 µm·cm, a 75 nm diameter wire would have a resistance of about 34 Ω/µm. The junction can contribute significantly to the overall resistance. In Ag NW network, the NW resistance was found to be 5 Ω/µm while the junction resistance was ~25 Ω.[4] Considering similar values between NW and junction, the equivalent resistance of two nanowires intersecting at 90° to each other will be ~315 Ω. Thus, 3 Ω is roughly equivalent to $10^2$ pairs of intersecting NWs in parallel. The high resistance (200 Ω) example therefore is equivalent to very few intersecting NW pairs. We note that, the junction resistance can vary widely.[5] Additionally, the networks studied in this work were prepared differently than the published intersecting NW junctions. This can also lead to different junction resistances.

The magnitude of the current pulse was 10 mA for the cases shown in Fig. 2(b), Fig. 3, Fig. 4; 500 µA for the cases shown in Fig. 2(c) ($\varphi=90°$ and $\varphi=45°$), 1 mA for the case shown in Fig. 2(c), $\varphi=0°$, 400 µA for the cases shown in Fig. 2(d), $\varphi=90°$ and $\varphi=0°$ and 800 µA for the cases shown in Fig. 2(d), $\varphi=30°$. Different current magnitudes were used for the current induced magnetization reversal experiment as mentioned in the text. The period of the current pulse was 83 ms in all cases. The duration of the pulse was 1 ms for the cases shown in Fig. 2(b), Fig. 2(d), Fig. 3, Fig. 4; 12 ms for the cases shown in Fig. 2(c) and 500 µs in Fig. S4. For all the MR curves presented in this paper, multiple measurements (up to 60) were averaged to reduce the noise level. Some heating induced drift was present in the high resistance case [Fig. 2(c)], limited to <0.05% during a single measurement, while it was negligible for the other measurements. Drifts were corrected by subtracting a linear slope from the data.

To measure MR-FORC, the sample was first saturated at 8 kOe. Next, the applied field was reduced to a given reversal field $H_R$ and the resistance was measured as the applied field $H$ was swept back to positive saturation, tracing out a single MR-FORC. This process was repeated for a series of decreasing reversal fields $H_R$ ranging from 3.0 kOe to -6.8 kOe at 0.1 kOe interval to a saturation field of 8.0 kOe, thus generating a family of FORCs. Five individual families of FORCs

# Supporting Information

were measured following this procedure. To obtain the FORC distribution shown in Fig. 4 (b), these five individual measurements were averaged to minimize the noise and a mixed second order derivative was taken. Gaussian smoothing was performed in between the steps to further minimize the noise. The five families of FORCs were examined individually to analyze the stochastic behavior shown in Fig. 4(e).

**Off-axis Electron Holography Imaging**

The nanowires were first separated from the polycarbonate membrane and the Cu backing electrode by submerging the sample in dichloromethane for 20-30 minutes. Thereafter, a liquid exchange process was performed to transfer the NWs from the dichloromethane to an isopropyl alcohol (IPA) solution. The NWs were then drop-casted on holey carbon coated Cu grids.

To image the magnetic structures in the isolated and interconnected NWs, off-axis electron holography was carried out by using a probe-aberration-corrected transmission electron microscope (FEI Titan G2 60-300) in Lorentz mode at 300 KV. The biprism, which is located in place of selected-area apertures, was biased to a positive voltage of 120 V. The application of the voltage caused the reference wave (passed through the vacuum) and the object wave (passed through the specimen) to interfere. Then a hologram is formed in the overlapping area. The spacing of the interference fringe is about 3.1 nm, so the spatial resolution is about 9 nm. By using the commercial digital reconstruction software (HoloWorks), the total phase shift $\phi$ can be recovered, as described below.

$$\phi(x,y) = \phi_E(x,y) + \phi_M(x,y) = C_E \int V(\mathbf{r})dz - \frac{e}{\hbar} \iint B_\perp(\mathbf{r})dxdz \qquad (1)$$

It is clear that both magnetic field $\phi_M(x,y)$ and inner electric field $\phi_E(x,y)$ contribute to the phase shift. To interpret the magnetic contribution of primary interest, we use a flipping method. The basic principle is changing the sign of the magnetic contribution by reversing the specimen. We acquired paired holograms on the front and back of the specimen and realigned two phase images using our own script, then the pure magnetic contribution can be extracted from half of the difference between two aligned phase images. Finally, we obtained the in-plane magnetic field distribution and the magnetic contour.

# Supporting Information

**MICROMAGNETIC SIMULATIONS OF INTERSECTION-MEDIATED DOMAIN WALL PROPAGATION**

To gain further insight into the microscopic magnetization reversal process, we have performed 3D micromagnetic simulations using Mumax3.[6] The magnetization dynamics was simulated in a simulation space of 2000×2000×100 nm³ and a constant cell size of 3.9×3.9×3.1 nm³ by solving the Landau-Lifshitz-Gilbert (LLG) equation:

$$\frac{\partial \vec{m}}{\delta t} = \left(\frac{-\gamma}{1+\alpha^2}\right)[\vec{m} \times \vec{B}_{eff} + \alpha\{\vec{m} \times (\vec{m} \times \vec{B}_{eff})\}] \quad (2)$$

where $\alpha$ is the Gilbert damping coefficient and $\gamma$ is the gyromagnetic ratio. $\vec{m}$ is the normalized magnetization vector ($\vec{M}/M_s$), $M_s$ is the saturation magnetization and $\vec{B}_{eff}$ is the effective magnetic field having the following components:

$$\vec{B}_{eff} = \vec{B}_{demag} + \vec{B}_{exchange} + \vec{B}_{anis} \quad (3)$$

Here, $\vec{B}_{demag}$ is the effective field due to demagnetization energy and $\vec{B}_{exchange}$ is the Heisenberg exchange interaction. The NWs are modeled as two intersecting cylinders with periodic boundary conditions to minimize edge effects, and with the following parameters: exchange stiffness = 20 pJ/m, Gilbert damping coefficient = 0.5, and saturation magnetization $M_s$ = 1.2×10⁶ A/m.

We modeled two nanowires intersecting at 45° to each other. The lumped circuit model for the MR of two intersecting wires based on micromagnetic simulations is shown in Fig. S1. The two wires were divided into 12 regions, where regions 1 to 4 are the non-intersecting wire segments (Fig. S1a). We further divided the intersection into 8 regions, with regions 5, 6, 9, 10 being the top part (z >0) and regions 7, 8, 11, 12 being the bottom part (z < 0). MR in each region was obtained by finding the angle between the total magnetization and the current in that region. The direction of the current was considered to be along the nanowire except for the intersection where the current was assumed to be the vector summation of the currents entering from the two branches as shown by the red arrows in Fig. S2 inset. The MR in each region was calculated separately by $R_i = cos^2(\theta_i) * w_i$, where $i$ is the region number, $\theta$ is the angle between the local current and the total magnetization averaged within each region, and $w$ is the resistance of each region. Previously, similar methodology was employed in artificial spin ice systems to analyze their magneto-transport behavior.[7]

# Supporting Information

The intersection area was considered to have a uniaxial anisotropy ($1.2\times10^6$ J/m$^3$). As an extreme case, we purposely set the anisotropy axis to be perpendicular to the nominal current direction in the intersection. To mimic experimental finding that the resistivity is much enhanced in confined geometries[8] and intersections[9], $w$ was increased by 15 times in the intersection compared to the non-intersecting segments. Finally, the equivalent MR of the whole geometry was simulated using the lumped circuit model by $R_{total} = (R_1 \parallel R_2) + R_{intsection} + (R_3 \parallel R_4)$, where $R_{intsection} = (R_5 \parallel R_6 \parallel R_7 \parallel R_8) + (R_9 \parallel R_{10} \parallel R_{11} \parallel R_{12})$ as shown in Fig. S1(b). A scaling factor of 0.05 was used while calculating the total change in MR percentage.[7]

Fig. S2 shows the simulated MR curve for φ=90° along a descending-field sweep. This equivalent MR model qualitatively reproduced the experimental observations such as multiple kinks and a prominent dip at a positive field in the descending MR branch. Starting from positive saturation, MR first increased as the moments started to align along the nanowire long axis due to its shape anisotropy. In the meantime, the moments in the intersection started to get pinned along the uniaxial anisotropy axis which is perpendicular to the current, leading to a peak and then a dip before reaching remanence. Fig. S2i shows that during pinning the moments in the intersection were mostly along the anisotropy direction and those in the wire were mostly aligned along the wire. As the field was further decreased, a jump in the MR signal was observed, corresponding to vortex-like domain wall formation near the intersection (Fig. S2ii). Subsequently, a second dip was observed as the vortex domain walls expanded in the wire while still being pinned by the intersection (Fig. S2iii). Afterwards, another MR jump emerged, corresponding to switching of one of the wires via DW propagation (Fig. S2iv). Fig. S2iv and S2v show the micromagnetic configuration where only one of the wires switched and the intersection moments were aligned perpendicular to the current direction. This state persisted until the other wire and the intersection switched at a more negative field when the MR signal jumped back up (Fig. S2vi). Thus, these simulations show that the intersection plays a critical role in the reversal dynamics and can lead to multiple kinks in the simulated MR signal, which is reflected in the experimentally measured MR.

# Supporting Information

## EXAMPLES OF MAGNETORESISTANCE MEASUREMENT ON FEW INTERCONNECTED NANOWIRES

In the main text, we have discussed magnetoresistance (MR) of a few interconnected nanowires which showed multiple kinks indicating domain wall (DW) pinning/depinning at the intersections [Fig. 2c]. Intersection geometry, such as location of intersection along the nanowire axis and degree of overlap, varies across the sample. Therefore, DW nucleation and depinning fields are also expected to vary. Few such cases are shown in Fig. S3. The case shown in Fig. S3(a) corresponds to a sample area with a resistance value ~ 60 Ω. While multiple MR kinks were present, similar to that shown in Fig. 2(c) of the main text, the number of switching steps and depinning fields were different. For $\varphi= 30°$, five distinct switching steps were observed. Similar behaviors were observed in multiple other samples. Three representative cases are shown in Fig. S3(b). These examples show that the intersection geometry is crucial in determining the DW pinning/depinning and consequently magnetization reversal behavior of the interconnected nanowires.

## CURRENT-DRIVEN DOMAIN WALL PROPAGATION THROUGH THE NETWORKS

The possibility of controlling DW nucleation and pinning/depinning at the intersection using current pulses was explored. During the field sweep, 500 µs-long pulses of different current densities were applied at each field. The results for $\varphi =35°$ are presented in Fig. S4(a). At lower currents (I=10 mA, 12 mA), some pinning and depinning events can be seen. However, when the current is increased, additional MR jumps appeared, indicating that the reversal of the DW state proceeded through a different route. Most likely, the applied current facilitated stabilization of DW at the intersection, which subsequently resulted in the switching of a section of the NW network. Several phenomena such as current induced spin torque, Oersted field or Joule heating could contribute to this behavior.[10-14] As the current is increased further, the peak position (i.e. DW depinning field) monotonically changed [Fig. S4(a)]. Similar trend was observed when the tilting angle was changed to $\varphi=90°$ albeit with a different slope of DW depinning field vs. applied current [Fig. S4(b)]. Thus, these results suggest that current pulses of a given magnitude can enable accessing different magnetic states, which leads to a new mode of magnetization reversal. The

# Supporting Information

possibility of controlling magnetization reversal using current could be a key ingredient for implementing memristor and racetrack devices using these networks.

## EXTENDED MR DATA IN A NETWORK WITH MANY NANOWIRES AND HIGH DENSITY OF INTERCONNECTS

In the main text we showed MR behavior for 9 different magnetic field angles in Fig. 3. Here, we provide enlarged view of measurements for 12 different field angles in Fig. S5.

## ANIMATION OF THE MR-FORCS

For first-order reversal curve (FORC) measurement, the applied field was reduced to a given reversal field $H_R$ after positive saturation. From this reversal field (i.e. a point on the descending branch of the major loop), the resistance was then measured as the applied field $H$ was swept back towards positive saturation, thereby tracing out a single MR-FORC. This process was repeated for a series of decreasing reversal fields, creating a family of FORCs. An animation of the full FORC family is provided in the Supplementary movie. Here, the sample location was the same as mentioned in Fig. 3 of the main text with a tilting angle $\varphi = 35°$. The MR-FORCs are overlaid on the major loop.

# Supporting Information

# Supporting Information

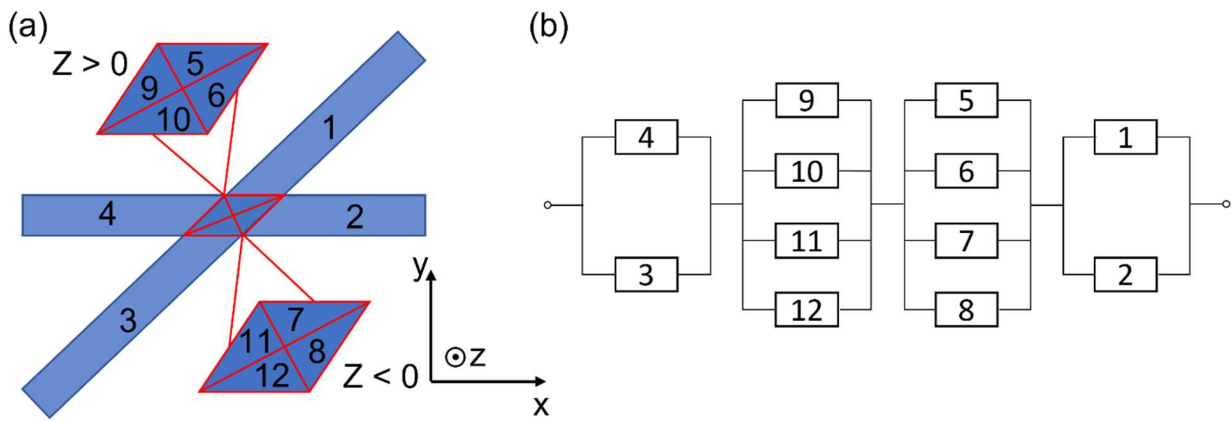

**Figure S1**. **Magnetoresistance calculation from micromagnetic simulation.** (a) Schematic showing the regions in the micromagnetic model. (b) Equivalent resistance model used to calculate the total MR.

**Supporting Information**

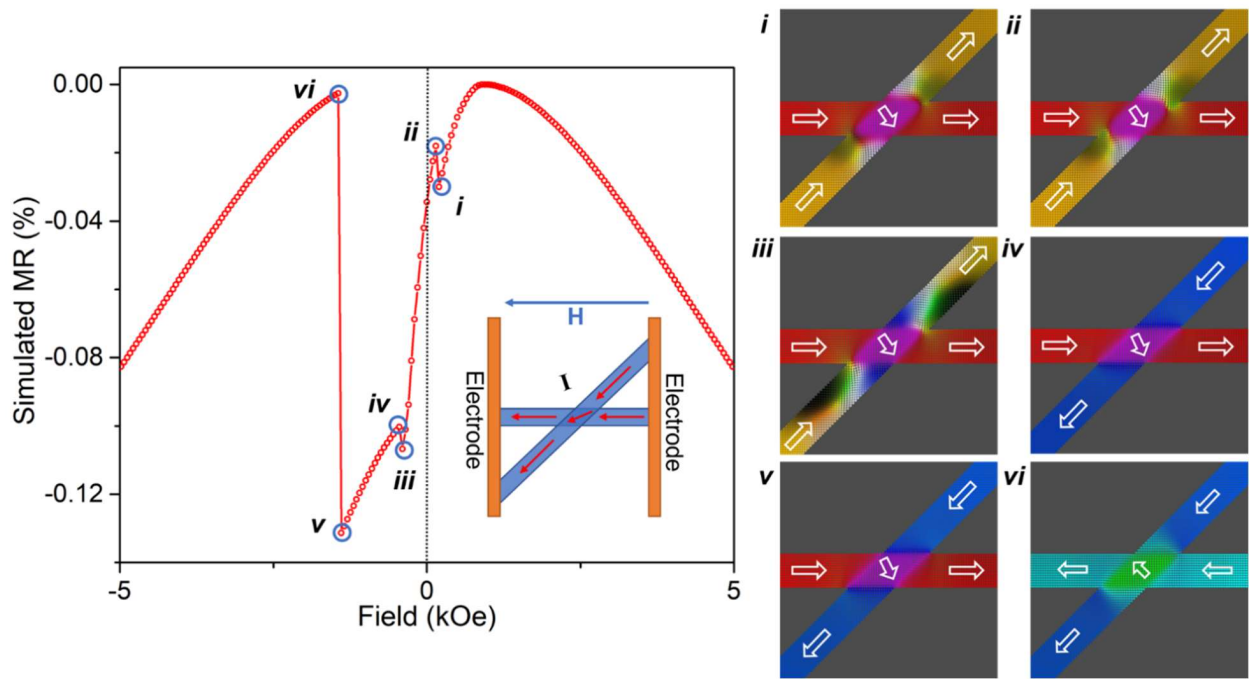

**Figure S2. Micromagnetic simulations.** Left panel shows the simulated MR curves when the field is swept from positive to negative saturation. In the inset the modeled geometry is shown. The red arrows indicate the assumed current direction. Multiple kinks can be seen in the MR curve similar to the experimental observation. (i-vi) show micromagnetic configuration close to these kinks, corresponding to the circles marked in the MR curve. The white arrows point to the general direction of the moments in each region.

**Supporting Information**

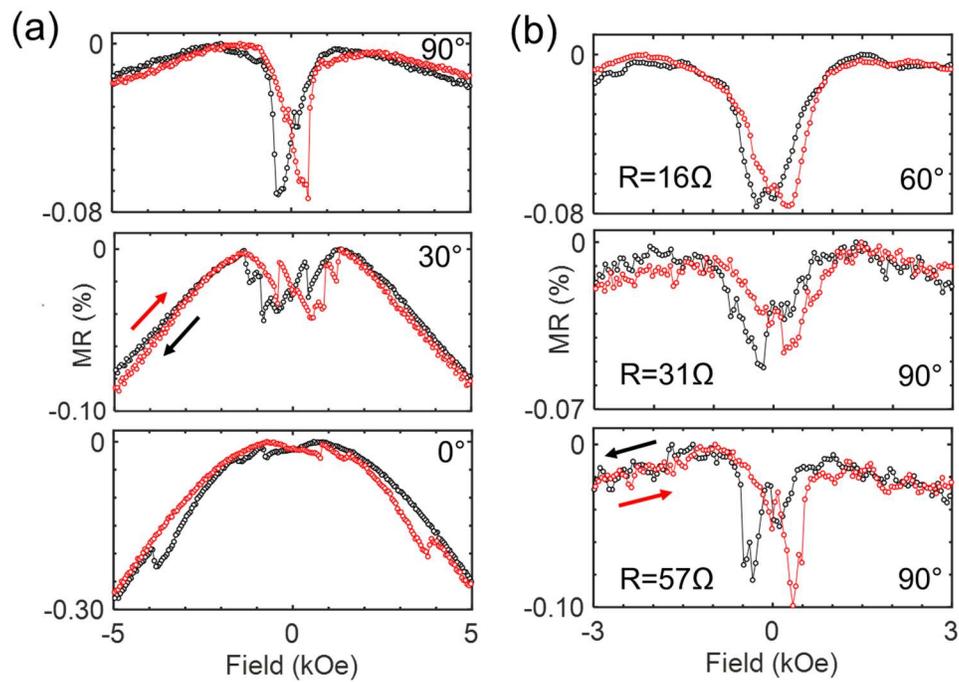

**Figure S3**. **Different examples of MR measurement on higher resistance networks.** (a) MR measurements performed with field applied at different angles relative to the ion-tracking plane normal. Resistance of the network ~60 Ω. (b) Examples of three other representative samples showing multiple kinks. The arrows indicate the sweep direction.

**Supporting Information**

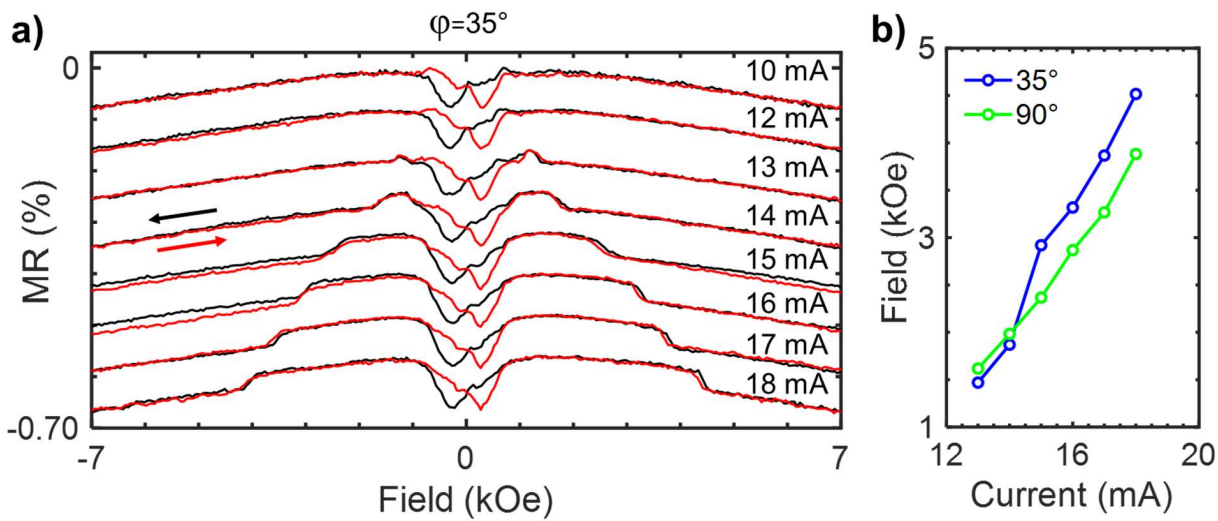

**Figure S4. Effect of current pulse on magnetization switching.** (a) MR curves showing shift of domain wall pinning with current density for φ=35°. The arrows show the sweep direction of magnetic field. (b) Pinning field vs. current density for two different magnetic field directions.

**Supporting Information**

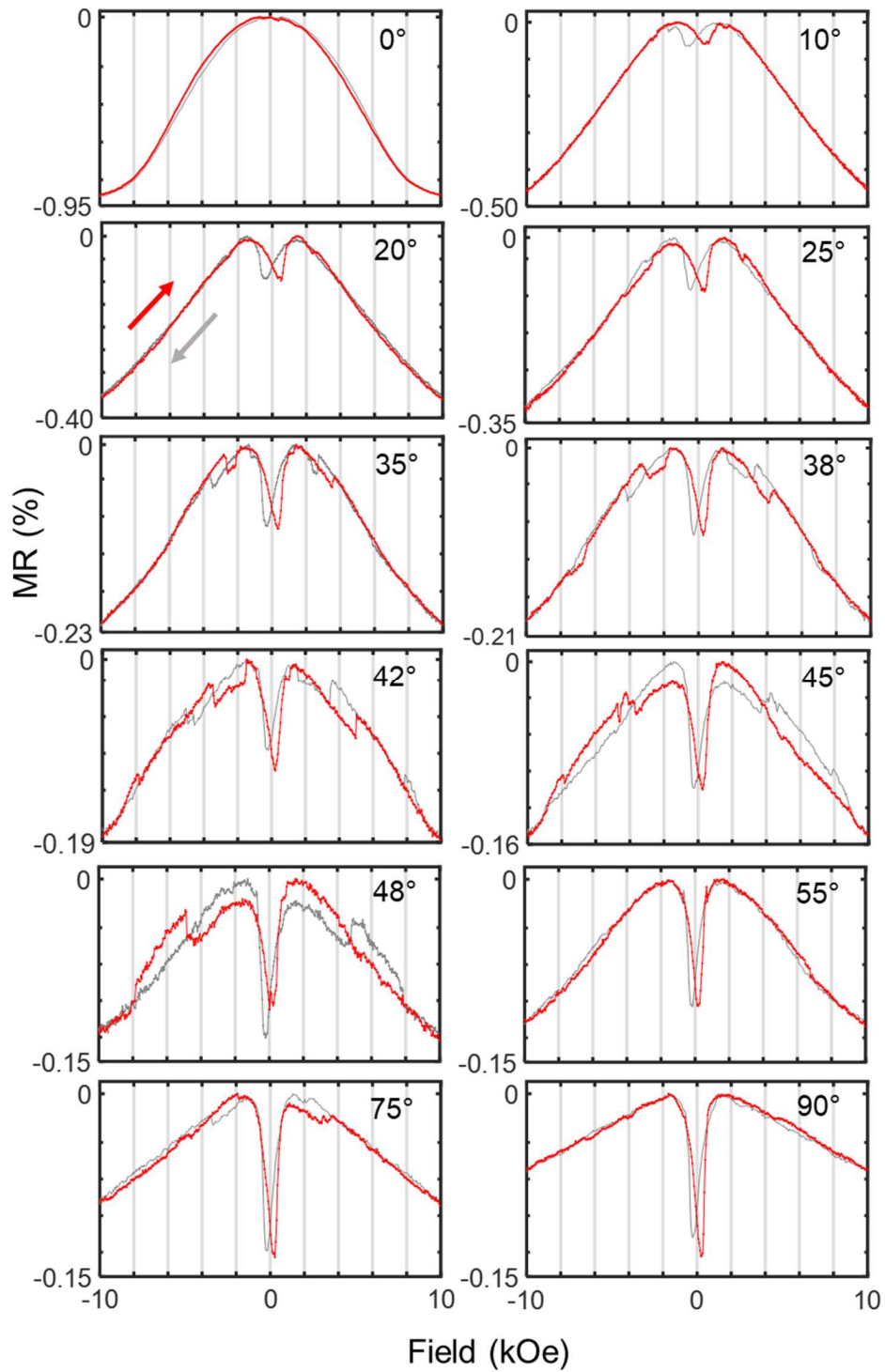

**Figure S5**. **MR behavior in a network with many nanowires and high density of interconnects.** MR curves for different applied magnetic field angle. For clarity, ascending-field sweep is shown in red, while descending-field sweep is shown in gray.

13